\documentclass[journal,10pt]{IEEEtran}
\usepackage{cite}
\ifCLASSINFOpdf
\else
  \usepackage[dvips]{graphicx}
  \graphicspath{{../eps/}}
  \DeclareGraphicsExtensions{.eps}
\fi
\usepackage{amsmath}
\usepackage{bm}
\usepackage{amssymb}
\usepackage{array}
\usepackage{xcolor}
\usepackage{multirow}
\usepackage{url}
\usepackage{lipsum}
\usepackage{mathtools}
\usepackage{cuted}
\usepackage{float}
    \usepackage[caption=false,font=footnotesize]{subfig}

\usepackage{stfloats}
\usepackage[numbers,sort,compress]{natbib}
\usepackage[linesnumbered,ruled,vlined]{algorithm2e}

\definecolor{PAPER_FRAGMENT}{RGB}{5,5,115}
\begin{document}

\title{Chirped DFT-s-OFDM: A new single-carrier waveform with enhanced LMMSE noise suppression}

\author{Yujie~Liu,~\IEEEmembership{Member,~IEEE,}
Yong~Liang~Guan,~\IEEEmembership{Senior Member,~IEEE,}
David~Gonz\'alez G.,~\IEEEmembership{Senior Member,~IEEE},
Halim~Yanikomeroglu,~\IEEEmembership{Fellow,~IEEE}

\thanks{This paper has been accepted by IEEE TVT. Copyright (c) 20xx IEEE. Personal use of this material is permitted. However, permission to use this material for any other purposes must be obtained from the IEEE by sending a request to pubs-permissions@ieee.org. This study is supported under the RIE2020 Industry Alignment Fund - Industry Collaboration Projects (IAF-ICP) Funding Initiative, as well as cash and in-kind contribution from the industry partner(s). (\emph{Corresponding author: Yujie Liu}.)}
 \thanks{Yujie Liu and Yong Liang Guan are with Nanyang Technological University, Singapore 637553. David Gonz\'alez G. is with Continental AG, 65936 Frankfurt am Main, Germany. Halim Yanikomeroglu is with Carleton University, Ottawa, ON K1S5B6, Canada (e-mail: yujie.liu@ieee.org; eylguan@ntu.edu.sg; david.gonzalez.g@ieee.org; halim@sce.carleton.ca).}}

\maketitle

\IEEEpeerreviewmaketitle

\begin{abstract}
In this paper, a new single-carrier waveform, called chirped discrete Fourier transform spread orthogonal frequency division multiplexing \mbox{(DFT-s-OFDM)}, is proposed for the sixth generation of communications. As the initial study on this waveform, its performance is analyzed and evaluated in single-user uplink multiple access communications, using a delay-Doppler channel model with independent and identically distributed path amplitudes and different integer delays. By chirping DFT-s-OFDM in the time domain, it maintains the low peak-to-average-power ratio of DFT-s-OFDM. Thanks to full-band transmission and symbols retransmission enabled by chirping and discrete Fourier transform precoding, it enhances noise suppression of linear minimum mean square error equalization. By using pairwise error probability analysis, the derived bit error rate upper bound is close to the simulated ones and the diversity order analysis confirms that it achieves full frequency diversity.
\end{abstract}

\section{Introduction}
Transmission waveform has been regarded as one of important components in every generation of \mbox{communications}. Orthogonal frequency division multiplexing (OFDM) was adopted for downlink transmissions in the fourth (4G) and fifth generations (5G) of \mbox{communications}. For the sake of low \mbox{peak-to-average-power-ratio (PAPR)}, discrete Fourier transform spread OFDM (DFT-s-OFDM) \cite{9264180,4099344,9143507,4085722} was selected for uplink transmissions. However, they may not be well-suited for the high-mobility applications in the sixth generation (6G) of communications. 6G waveform design has thus attracted significant attention from both academia and industry recently.

The 6G waveform candidates in the literature include orthogonal time frequency space (OTFS) \cite{9741716,10050811,10250854}, orthogonal delay-Doppler division multiplexing (ODDM) \cite{oddm}, orthogonal time sequency multiplexing (OTSM) \cite{otsm}, frequency-modulated OFDM (FM-OFDM) \cite{fmofdm}, affine frequency division multiplexing (AFDM) \cite{afdm_twc}, etc. However, OTFS, ODDM and OTSM all have high PAPR. PAPR reduction \cite{9088986} has been studied for OTFS. However, it introduces clipping noise and additional compensation methods are required \cite{9345729}. FM-OFDM is a variant of constant-envelope OFDM (CE-OFDM) \cite{4600180} and inherits its perfect PAPR. It performs phase accumulation of OFDM symbols before phase modulator and exhibits higher resilience to Doppler frequency than CE-OFDM \cite{fmofdm}. \mbox{However}, its benefits come at the expense of expanding bandwidth. Similarly to orthogonal chirping division multiplexing \cite{7523229}, AFDM adds chirping in the frequency and time domain before and after OFDM modulation, respectively. To cater for high-mobility communications, its chirp rates are properly selected according to maximum Doppler frequency. However, AFDM, as a variation of OFDM, retains its high PAPR.

In this correspondence, a new single-carrier waveform, called chirped DFT-s-OFDM, is proposed for 6G communications, with an initial study conducted in single-user uplink multiple access communications (MAC) using the delay-Doppler channel model \cite{afdm_twc} with independent and identically distributed (\emph{i.i.d.}) path amplitudes and different integer delays. The contributions of this paper are summarized as follow:
\begin{itemize}
\item To the best of the authors' knowledge, this is the first single-carrier chirping waveform featuring low \mbox{PAPR} for 6G communications. In contrast, the existing 6G waveform candidates, including OTFS \cite{9741716,10050811,10250854}, ODDM \cite{oddm}, OTSM \cite{otsm}, and AFDM \cite{afdm_twc}, suffer high PAPR.
    \item By using chirping and discrete Fourier transform (DFT) precoding, the proposed waveform enables full-band transmission and symbols retransmission, leading to enhanced noise suppression of linear minimum mean square error (LMMSE) equalization. Simulation results confirm that it outperforms DFT-s-OFDM, AFDM, OTFS, and OFDM in terms of bit error rate (BER) and output signal-to-noise-ratio (SNR) in single-user uplink MAC.
\item The BER upper bound and diversity order of proposed waveform are derived based on pairwise error probability (PEP). The simulated BER closely approaches the derived BER upper bound, and the derived diversity order corresponds to the number of integer delays, confirming full frequency diversity over delay-Doppler channels with \emph{i.i.d.} path amplitudes and different integer delays.
\end{itemize}

This correspondence is organized as follows. The proposed chirped DFT-s-OFDM is described in Section II, followed by its performance analysis in Section III. Sections~IV and V provide simulation results and conclusion.

\begin{figure*}[htbp]
	\centering
	\includegraphics[width=18cm]{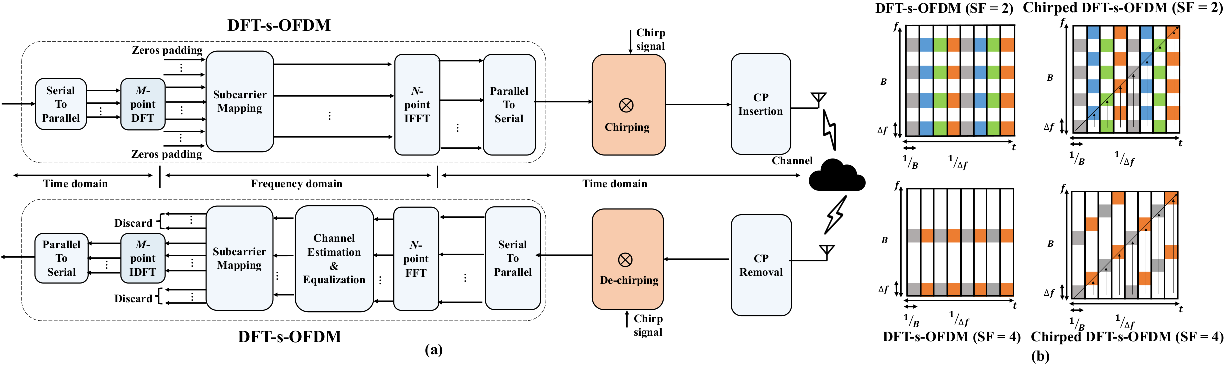}
	\caption{a) Block diagram of proposed chirped DFT-s-OFDM, and b) time-frequency diagrams of (chirped) DFT-s-OFDM, with $N=8$ and $c_{\rm{r}}=1/8$.}
	\label{Fig.dft_chirp_s_ofdm}
\end{figure*}

\section{System Model}

\subsection{Modulation and Demodulation}
Considering single-user uplink MAC, the block diagram of proposed chirped DFT-s-OFDM is illustrated in Fig. \ref{Fig.dft_chirp_s_ofdm}a. 
At transmitter side, it includes two main blocks, with DFT-s-OFDM modulation followed by chirping. The un-modulated time-domain data vector of length $M$ is denoted as $\mathbf{x}=[x[0],x[1],\cdots,x[M-1]]^T$. {The elements of data vector should be \emph{i.i.d.}, \emph{i.e.}, $\mathbb{E}\{\mathbf{x}\mathbf{x}^H\}=\mathbf{I}_M$, where each data symbol has unit power and $\mathbf{I}_M$ is an identity matrix of size $M$.}
After $M$-point DFT, the frequency-domain data vector of length $M$ is given by
$\mathbf{X}=\mathbf{F}_M\mathbf{x}$,
 with $\mathbf{F}_M$ being $M$-point DFT matrix. After subcarrier mapping, the data vector of length $N$ is written as
$ \mathbf{S}=\mathbf{P}\mathbf{X}$,
  with $\mathbf{P}$ being a subcarrier mapping matrix of size $N\times M$ ($N>M$). The DFT spreading factor is defined as $\mathrm{SF}=N/M$ with $\mathrm{SF}>1$. For the sake of low PAPR, interleaved subcarrier mapping with an \emph{integer} value of $\mathrm{SF}$ \cite{4099344,9143507,4085722} is considered in this paper, yielding $\mathbf{P}=\mathbf{I}_{M}\otimes [1;\mathbf{0}_{(\mathrm{SF}-1)\times 1}]$. Inverse fast Fourier transform~(FFT) is then implemented, resulting in time-domain signal
 $ \mathbf{s}=\mathbf{F}_N^H\mathbf{S}$,
   with $\mathbf{F}_N$ being $N$-point FFT matrix. The DFT-s-OFDM signal $\mathbf{s}$ is then chirped with a linear chirp signal $\mathbf{c}$ of chirp rate $c_{\rm{r}}$, obtaining the chirped DFT-s-OFDM signal denoted as $\mathbf{s}_{\mathrm{c}}$ as follow:
   \begin{equation}
   \mathbf{s}_{\mathrm{c}}=\mathrm{diag}\{\mathbf{c}\}\mathbf{s}=\mathbf{C}\mathbf{s}=\mathbf{C}\mathbf{F}_N^H\mathbf{P}\mathbf{F}_M\mathbf{x},
   \end{equation}
    with $\mathbf{c}=[1,e^{j\pi c_{\rm{r}}},\cdots,e^{j\pi c_{\rm{r}}n^2},\cdots,e^{j\pi c_{\rm{r}}(N-1)^2}]$. 

A cyclic prefix of length $L_{\mathrm{cp}}$ ($L_{\mathrm{cp}}\geq L-1$) is appended to $\mathbf{s}_{\mathrm{c}}$ before transmission, with $L$ being the number of resolvable paths. $h[n,l]$ is defined as channel gain of $l$-th channel path at $n$-th time instant, with $l=1,2,\cdots,L$ and $n=0,1,\cdots,N-1$. Define $\bar{f}_{\textrm{max}}=\frac{f_{\textrm{c}}v}{c}/\Delta f$ as the normalized maximum Doppler frequency with respect to subcarrier spacing, with $f_{\textrm{c}}$, $v$, $c$, and $\Delta f$ being carrier frequency, velocity, speed of light, and subcarrier spacing, respectively. The delay-Doppler channel model in \cite{afdm_twc} is used in this paper:
\begin{equation}
h[n,l]=\sum\nolimits_{p=1}^{L}h_pe^{j2\pi \nu_pn/N}\delta(l-l_p),
\label{h_n_l}
\end{equation}
with $h_p$, $\nu_p$, and $l_p$ being the channel amplitude, normalized Doppler shift, and integer delay of $p$-th path. $h_p$ is modeled as an \emph{i.i.d.} complex Gaussian random variable with zero mean and variance of $1/L$ and $\nu_p$ is randomly selected from $-\bar{f}_{\textrm{max}}$ to $\bar{f}_{\textrm{max}}$ \cite{afdm_twc}. This paper assumes that different paths have different integer delays. However, \eqref{h_n_l} can be extended to the scenarios with multiple paths sharing the identical integer delay by setting $l_p=l_q$ for $p$-th and $q$-th path, while with $h_p\neq h_q$ and $\nu_p\neq \nu_q$ \cite{afdm_twc}. The time-domain channel matrix is expressed as
$\mathbf{H}_{\rm{t}}=\sum\nolimits_{l=1}^{L}h_l\mathbf{D}_l\bm{\Pi}^{l}$,
where $\mathbf{D}_l=\mathrm{diag}\{1,\cdots,e^{j2\pi \nu_ln/N},\cdots,e^{j2\pi \nu_l(N-1)/N}\}$ and $\bm{\Pi}$ is the forward cyclic-shift matrix defined as (25) of \cite{afdm_twc}. 



The received time-domain signal vector of length $N$ after the removal of cyclic prefix is expressed as
\begin{equation}
\mathbf{r}=\mathbf{H}_{\mathrm{t}}\mathbf{C}\mathbf{F}_N^H\mathbf{P}\mathbf{F}_M\mathbf{x}+\mathbf{w},
\end{equation}
with $\textbf{w}$ being an additive white Gaussian noise vector of variance $\sigma^2$. Define $\mathbf{z}=\mathbf{F}_N\mathbf{C}^H\mathbf{w}$. After dechirping and $N$-point FFT, the frequency-domain received signal vector is 
\begin{align}
\mathbf{y}&=\mathbf{F}_N\mathbf{C}^H\textbf{H}_{\mathrm{t}}\mathbf{C}\mathbf{F}_N^H\mathbf{P}\mathbf{F}_M\mathbf{x}+\mathbf{z}.
\label{eq_sm}
\end{align}

\subsection{Time-Frequency Diagram}
The time-frequency diagrams of (chirped) DFT-s-OFDM are illustrated in Fig. \ref{Fig.dft_chirp_s_ofdm}b for a certain user, with $N=8$ and $c_{\rm{r}}=1/8$. Regarding DFT-s-OFDM with \mbox{$\rm{SF}=2$}, information symbols are mapped in the time domain in four different colors and then are repeated in the time domain with reduced amplitude, with each symbol occupying the assigned subcarrier bands \cite{4099344,9143507,4085722}. The un-assigned subcarrier bands are indicated in white color and can be used by other users. {A linear chirp signal with chirp rate $c_{\rm{r}}=1/N$ is used, implying that the subcarrier frequency is increased by one subcarrier spacing per time instant. The original subcarrier $k$ ($k=0,1,\cdots,N-1$) at $n=0$ would be changed to $k+n$ at time instant $n$.  For example, the subcarrier 0 remains at subcarrier $0$ at $n=0$, hops to subcarrier $1$ at $n=1$, etc.} After chirping, the data symbols, \emph{e.g.}, in blue and orange color, can hop to the un-assigned subcarrier bands to enable full band transmission. {It then provides more observations than unknowns, and the redundant information from more equations can thus enhance noise suppression in the frequency domain.} Besides, when $\rm{SF}$ increases to $4$ and each symbol would be repeated $4$ times in the time domain, additional noise suppression can be obtained.

Note that chirp signal was applied with \mbox{DFT-s-OFDM} in~\cite{9264180}, which however is used in the frequency domain and at transmitter only. It also presents limited performance gains over DFT-s-OFDM \cite{9264180}. In contrast, chirp signal is used in the time domain at both transmitter and receiver for the proposed waveform, referred to as chirping and dechirping in Fig. \ref{Fig.dft_chirp_s_ofdm}a. According to \cite{afdm_twc}, chirping in the time domain is expected to have a greater impact than that in the frequency domain.

\section{Performance Analysis}

\subsection{Error Performance Analysis Using ML Equalizer}
{This subsection aims to apply PEP analysis technique of maximum likelihood (ML) equalizer to derive the BER upper bound and diversity order of proposed chirped DFT-s-OFDM.
The system model in \eqref{eq_sm} is rewritten as
\begin{align}
\mathbf{y}=\sum\nolimits_{l=1}^L\mathbf{F}_N\mathbf{C}^H\mathbf{D}_l\bm{\Pi}^{l}\mathbf{C}\mathbf{F}_N^H\mathbf{P}\mathbf{F}_M\mathbf{x}h_l+\mathbf{z}.
\label{eq_y_ml}
\end{align}
Denote $\mathbf{E}_l(\mathbf{x})=\mathbf{F}_N\mathbf{C}^H\mathbf{D}_l\bm{\Pi}^{l}\mathbf{C}\mathbf{F}_N^H\mathbf{P}\mathbf{F}_M\mathbf{x}$. $\mathbf{y}$ is expressed as
\begin{align}
\mathbf{y}=\mathbf{E}(\mathbf{x})\mathbf{h}+\mathbf{z},
\end{align}
with $\mathbf{E}=[\mathbf{E}_1,\mathbf{E}_2,\cdots,\mathbf{E}_{L}]$ and $\mathbf{h}=[h_1,h_2,\cdots,h_L]^T$. 
{$\mathbf{x}$ can be estimated by using ML equalizer:
\begin{equation}
\mathbf{\hat{x}}=\min_{\mathbf{\tilde{x}}\in\mathbb{K}}\lVert \mathbf{y}- \mathbf{E}(\mathbf{\tilde{x}})\mathbf{h}\rVert,
\end{equation}
where $\mathbf{\tilde{x}}$ denotes the sequence candidate of data symbols and each element is chosen from a certain modulation alphabet $\mathbb{K}$. ML equalizer would incur high complexity, making it less suitable for big data vector size ($M$) and big symbol alphabet size ($Q$). Nevertheless, persurvivor processing (PSP) can be used to approximate ML algorithm with low complexity \cite{9036080}.}

Denote $\{\mathbf{x}\rightarrow \hat{\mathbf{x}}\}$ as the pairwise error event, where $\mathbf{x}$ represents the transmitted signal vector and $\hat{\mathbf{x}}$ the erroneously detected signal vector using ML equalizer. $\mathbf{x}$ and $\hat{\mathbf{x}}$ are chosen from a certain modulation alphabet, \emph{i.e.}, $\mathbf{x}\in\mathbb{K}$ and $\hat{\mathbf{x}}\in\mathbb{K}$, but with $\mathbf{x}\neq\hat{\mathbf{x}}$. Define $\bm{\Theta}(\mathbf{x},\hat{\mathbf{x}})=(\mathbf{E}(\mathbf{x})-\mathbf{E}(\hat{\mathbf{x}}))^H(\mathbf{E}(\mathbf{x})-\mathbf{E}(\hat{\mathbf{x}}))$. The rank of $\bm{\Theta}(\mathbf{x},\hat{\mathbf{x}})$ and its non-zero eigenvalues are denoted as $R$ and $\{\lambda_1,\lambda_2,\cdots,\lambda_R\}$. The SNR for each data symbol is denoted as $\gamma$, \emph{i.e.,} $\gamma=\frac{1}{\sigma^2}$. Following the derivation in \cite{10050811} and \cite{10250854}, the PEP can be expressed as
\begin{small}
\begin{equation}
P_E(\mathbf{x}\rightarrow \hat{\mathbf{x}})\approx\frac{1}{12}\prod\nolimits_{r=1}^R\frac{1}{1+\frac{\gamma\lambda_r}{4L}}+\frac{1}{4}\prod\nolimits_{r=1}^R\frac{1}{1+\frac{\gamma\lambda_r}{3L}}.
\label{eq_pep}
\end{equation}
\end{small}
At high SNR, \eqref{eq_pep} can be approximated as
\begin{small}
\begin{align}
P_E(\mathbf{x}\rightarrow \hat{\mathbf{x}})\approx \frac{1}{12}\left [\left(\prod\nolimits_{r=1}^R\lambda_r\right)^{\frac{1}{R}}\frac{\gamma}{4L}\right]^{-R}+\notag\nonumber\\
\frac{1}{4}\left[\left(\prod\nolimits_{r=1}^R\lambda_r\right)^{\frac{1}{R}}\frac{\gamma}{3L}\right]^{-R}.
\end{align}
\end{small}
The average BER can then be obtained by using the union bound technique \cite{10050811,10250854} as follow:
\begin{small}
\begin{equation}
P_e\leq \frac{1}{Q^MM\rm{log}_2Q}\sum\nolimits_{\mathbf{x}\in\mathbb{K}}\sum\nolimits_{\hat{\mathbf{x}}\in\mathbb{K},\mathbf{x}\neq\hat{\mathbf{x}}}P_E(\mathbf{x}\rightarrow \hat{\mathbf{x}})d(\mathbf{x},\hat{\mathbf{x}}),
\label{Pe}
\end{equation}
\end{small}
with $d(\mathbf{x},\hat{\mathbf{x}})$ being the number of bits in difference between $\mathbf{x}$ and $\hat{\mathbf{x}}$. The diversity order of chirped DFT-s-OFDM is given~by
\begin{equation}
G_D=\min_{\mathbf{x}\in\mathbb{K},\hat{\mathbf{x}}\in\mathbb{K},\mathbf{x}\neq\hat{\mathbf{x}}}\rm{rank}(\mathbf{\Theta}(\mathbf{x},\hat{\mathbf{x}})).
\label{eq_do}
\end{equation}
{With $L=3$, Table \ref{tab1} shows the diversity order $G_D$ of (chirped) DFT-s-OFDM computed using \eqref{eq_do} with different combinations of Doppler shifts. The diversity order for DFT-s-OFDM is calculated by substituting $c_{\rm{r}}=0$ to \eqref{eq_do}. It can be seen that DFT-s-OFDM cannot achieve full frequency diversity and its diversity order varies with Doppler shifts. In contrast, the proposed waveform can exploit full frequency diversity and shows high resilience to Doppler shifts.} {To enable full frequency diversity, the chirp rate should be chosen from $c_{\rm{r}}=(a+b\times \mathrm{SF})/N$ to enable full bandwidth transmission, with $a=1,3,\cdots,\rm{SF}-1$ and $b$ being zero or positive integer value. For example, with $\mathrm{SF}=4$, $c_{\rm{r}}$ should be chosen from $\{1/N, 3/N, 5/N, ...\}$.

\begin{table}[!t]
	\centering
	\caption{Diversity order $G_D$ of (chirped) DFT-s-OFDM with $L=3$ calculated using \eqref{eq_do}. Note that for D: $v_0$, $v_1$, and $v_2$ are randomly generated from $-\bar{f}_{\textrm{max}}$ to $\bar{f}_{\textrm{max}}$ and the diversity order is calculated by averaging over $100,000$ realizations.}
	\begin{tabular}{|c|c|c|}
		\hline
Doppler shifts & \multirow{ 2}{*}{DFT-s-OFDM} & Proposed  \\
$[v_0,v_1,v_2]$ &  & chirped DFT-s-OFDM \\\hline
A: $[0,0,0]$ & $1$ & $3$\\\hline
B: $[-\bar{f}_{\textrm{max}},0,\bar{f}_{\textrm{max}}]$ &$2$ &$3$ \\\hline
C: $[-\bar{f}_{\textrm{max}},-\bar{f}_{\textrm{max}},\bar{f}_{\textrm{max}}]$&$2$ &$3$ \\\hline
D: Random &$1.9$ &$3$ \\\hline
	\end{tabular}
\label{tab1}
\end{table}

\begin{figure}[!t]
	\centering
	\includegraphics[width = 0.5\textwidth]{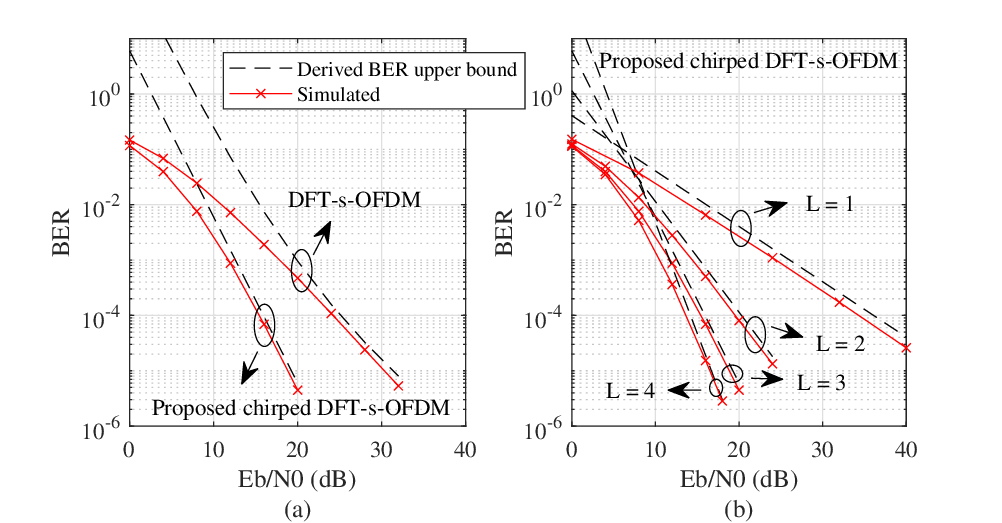}
	\caption{Derived BER upper bound \eqref{Pe} and simulated BER of (a) (chirped) DFT-s-OFDM with $L=3$, and (b) proposed chirped DFT-s-OFDM.}
	\label{Fig.ml}
\end{figure}

Fig. \ref{Fig.ml} shows the derived BER upper bound~\eqref{Pe} and simulated BER of (chirped) DFT-s-OFDM using ML equalizer. The DFT size and IFFT size are $M=2$ and $N=8$. The maximum Doppler frequency, velocity, and subcarrier spacing are $f_{\rm{max}}=2\,\rm{KHz}$, $v=500\,\rm{km/h}$, and $\Delta f=15\,\rm{KHz}$. Binary phase shift keying (BPSK) modulation is adopted. According to Fig. \ref{Fig.ml}, the simulated BER is close to the derived BER upper bound especially at high SNRs. The BER curves for DFT-s-OFDM in Fig. 2a exhibit a flatter slope, suggesting that DFT-s-OFDM has a lower diversity order than chirped DFT-s-OFDM. For chirped DFT-s-OFDM, its diversity order calculated from its BER curvers is $L$ in Fig. \ref{Fig.ml}b. 

The above analysis of proposed waveform assumes all paths have different integer delays and \emph{i.i.d.} path amplitudes. When some paths share the same integer delays, in addition to frequency diversity, Doppler/time diversity could be exploited thanks to Doppler spread \cite{9741716} and a detailed analysis will be addressed in future work. If path amplitudes are correlated (not \emph{i.i.d.}), full frequency diversity might not be attained.

\begin{figure*}[htbp]
	\centering
	\subfloat[Output signal power ($\mathrm{diag}\{\mathbf{H}_{\rm{eff}}^H\mathbf{G}^H\mathbf{G}\mathbf{H}_{\rm{eff}}\}$).]
	{\label{Fig:signal}
		\includegraphics[width = 0.33\textwidth]{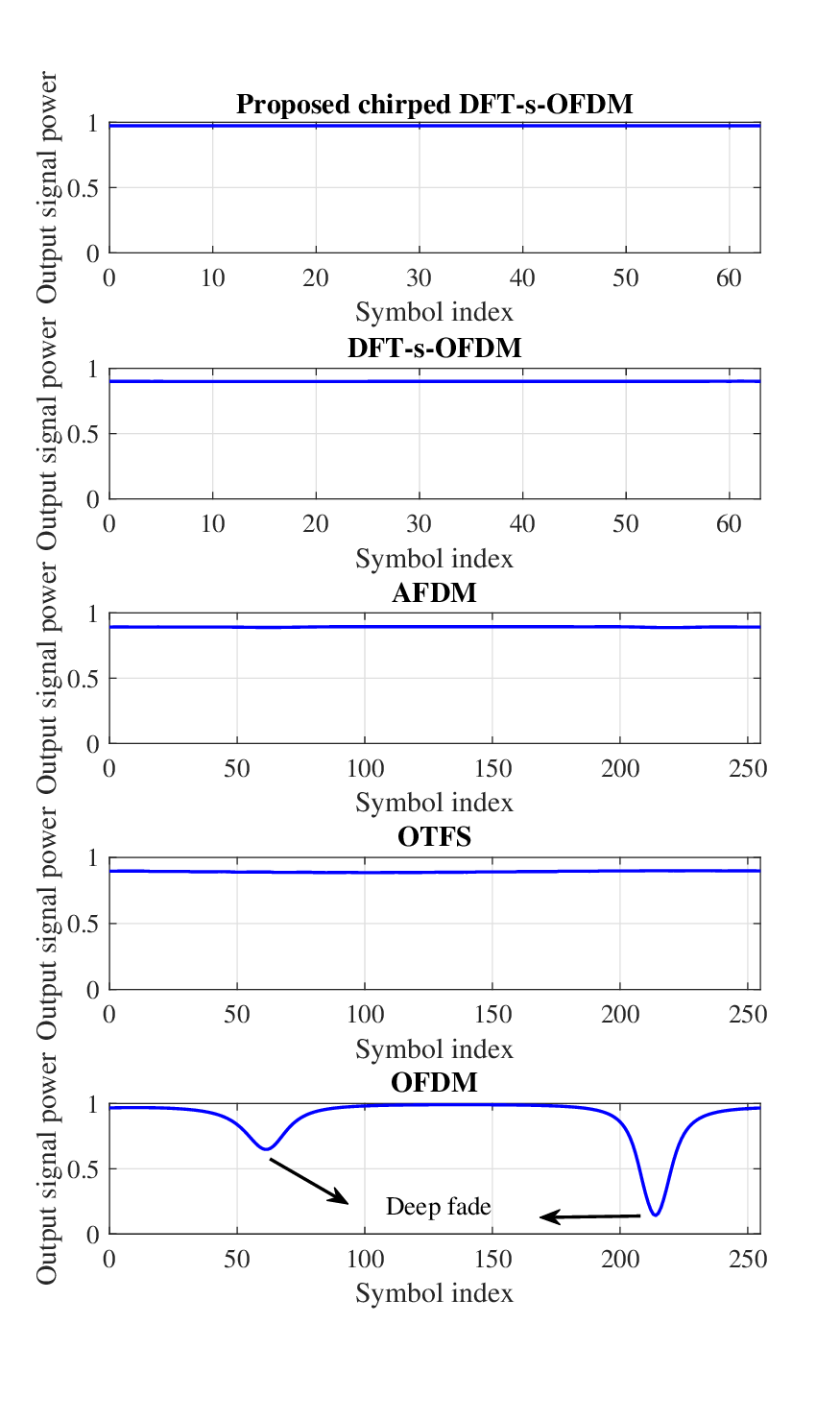}}		
	\subfloat[Output noise power $(\sigma^2\mathrm{diag}\{\mathbf{G}\mathbf{G}^H)\}$.]
	{\label{Fig:noise}
		\includegraphics[width = 0.33\textwidth]{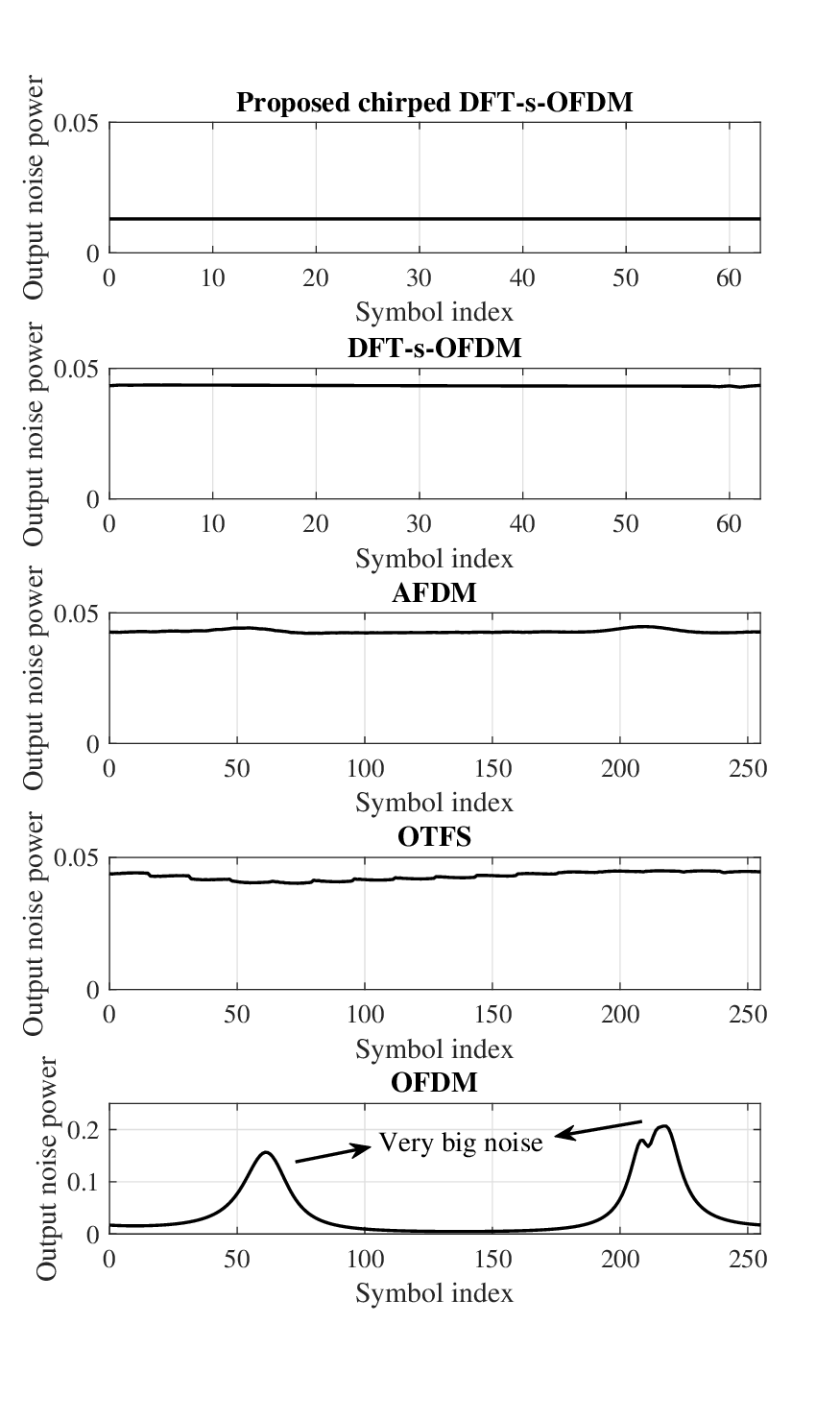}}		
	\subfloat[Output SNR (the ratio of Figs. \ref{Fig:signal} and \ref{Fig:noise}).]
	{\label{Fig:output_snr}
		\includegraphics[width = 0.33\textwidth]{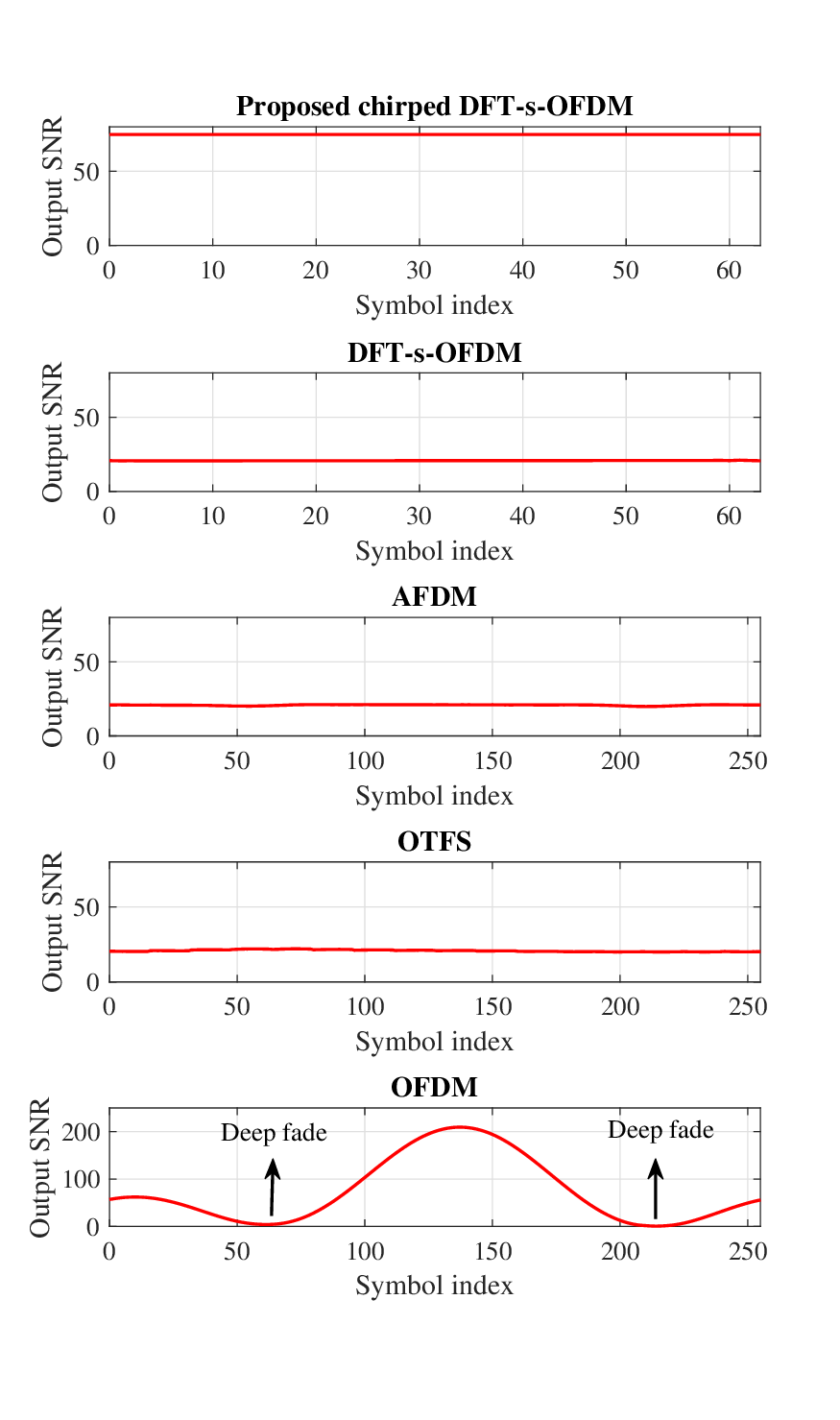}}		
	\caption{Output SNR analysis using LMMSE equalizer of proposed chirped DFT-s-OFDM, DFT-s-OFDM, AFDM, OTFS, and OFDM.}
	\label{Fig.output_snr_all}
\end{figure*}

\subsection{Output SNR Analysis Using LMMSE Equalizer}
This subsection aims to analyze output SNR of proposed chirped DFT-s-OFDM using LMMSE equalizer.
By letting $\mathbf{H}_{\rm{eff}}=\mathbf{F}_N\mathbf{C}^H\textbf{H}_{\mathrm{t}}\mathbf{C}\mathbf{F}_N^H\mathbf{P}\mathbf{F}_M$, \eqref{eq_sm} is rewritten as
\begin{align}
\mathbf{y}=\mathbf{H}_{\rm{eff}}\mathbf{x}+\mathbf{z}.
\end{align}
$\mathbf{x}$ can be estimated using LMMSE equalizer as
{\footnotesize
\begin{align}
&\hat{\mathbf{x}}_{\rm{LMMSE}}=\mathbf{H}_{\rm{eff}}^H\left(\mathbf{H}_{\rm{eff}}\mathbf{H}_{\rm{eff}}^H+\mathbf{I}_{N}\sigma^2\right)^{-1}\mathbf{y}=\mathbf{G}\mathbf{H}_{\rm{eff}}\mathbf{x}+\mathbf{G}\mathbf{z},
\label{x_lmmse}
\end{align}}
with $\mathbf{G}=\mathbf{H}_{\rm{eff}}^H\left(\mathbf{H}_{\rm{eff}}\mathbf{H}_{\rm{eff}}^H+\mathbf{I}_{N}\sigma^2\right)^{-1}$.
Note that intersymbol interference (ISI) has been mitigated to some extent through LMMSE equalizer. According to \eqref{x_lmmse}, without considering the impact of noise, the signal estimate is expressed as $\mathbf{G}\mathbf{H}_{\rm{eff}}\mathbf{x}$. The impact of ISI can be viewed from the non-diagonal elements of $\mathbf{G}\mathbf{H}_{\rm{eff}}$, which are small and negligible.

According to \cite{5730587}, the output SNR of chirped DFT-s-OFDM using LMMSE equalizer can be expressed as
\begin{align}
\textrm{Output\,SNR}=\frac{\mathbb{E}\{\mathbf{x}^H\mathbf{H}_{\rm{eff}}^H\mathbf{G}^H\mathbf{G}\mathbf{H}_{\rm{eff}}\mathbf{x}\}}{\mathbb{E}\{\mathbf{z}^H\mathbf{G}^H\mathbf{G}\mathbf{z}\}},
\label{output_snr}
\end{align}
where $\mathbb{E}(\cdot)$ indicates the expectation operation.
Due to the properties of trace operation, the following can be obtained: $\mathbf{x}^H\mathbf{H}_{\rm{eff}}^H\mathbf{G}^H\mathbf{G}\mathbf{H}_{\rm{eff}}\mathbf{x}=\rm{Tr}\{\mathbf{x}^H\mathbf{H}_{\rm{eff}}^H\mathbf{G}^H\mathbf{G}\mathbf{H}_{\rm{eff}}\mathbf{x}\}=\rm{Tr}\{\mathbf{x}\mathbf{x}^H\mathbf{H}_{\rm{eff}}^H\mathbf{G}^H\mathbf{G}\mathbf{H}_{\rm{eff}}\}$
and $\mathbf{z}^H\mathbf{G}^H\mathbf{G}\mathbf{z}=\rm{Tr}\{\mathbf{z}^H\mathbf{G}^H\mathbf{G}\mathbf{z}\}=\rm{Tr}\{\mathbf{z}\mathbf{z}^H\mathbf{G}^H\mathbf{G}\}$.
\eqref{output_snr} can be rewritten as
{\footnotesize
\begin{align}
&\textrm{Output\,SNR}=\frac{\mathbb{E}\{\rm{Tr}\{\mathbf{x}\mathbf{x}^H\mathbf{H}_{\rm{eff}}^H\mathbf{G}^H\mathbf{G}\mathbf{H}_{\rm{eff}}\}\}}{\mathbb{E}\{\rm{Tr}\{\mathbf{z}\mathbf{z}^H\mathbf{G}^H\mathbf{G}\}\}}\notag\nonumber\\
&=\frac{\rm{Tr}\{\mathbb{E}\{\mathbf{x}\mathbf{x}^H\}\mathbb{E}\{\mathbf{H}_{\rm{eff}}^H\mathbf{G}^H\mathbf{G}\mathbf{H}_{\rm{eff}}\}\}}{\rm{Tr}\{\mathbb{E}\{\mathbf{z}\mathbf{z}^H\}\mathbb{E}\{\mathbf{G}^H\mathbf{G}\}\}}
=\frac{\mathbb{E}\{\rm{Tr}\{\mathbf{H}_{\rm{eff}}^H\mathbf{G}^H\mathbf{G}\mathbf{H}_{\rm{eff}}\}\}}{\mathbb{E}\{\rm{Tr}\{\sigma^2\mathbf{G}\mathbf{G}^H\}\}},
\label{output_snr_1}
\end{align}}
with $\mathbb{E}\{\mathbf{x}\mathbf{x}^H\}=\mathbf{I}_{M}$, $\mathbb{E}\{\mathbf{z}\mathbf{z}^H\}=\sigma^2\mathbf{I}_N$, and $\rm{Tr}\{\mathbf{G}^H\mathbf{G}\}=\rm{Tr}\{\mathbf{G}\mathbf{G}^H\}$. 
As a result, the output signal power and output noise power are the sum of diagonal elements of $\mathbf{H}_{\rm{eff}}^H\mathbf{G}^H\mathbf{G}\mathbf{H}_{\rm{eff}}$ and $\sigma^2\mathbf{G}\mathbf{G}^H$, respectively. The $m$-th diagonal element of $\mathbf{H}_{\rm{eff}}^H\mathbf{G}^H\mathbf{G}\mathbf{H}_{\rm{eff}}$ and $\sigma^2\mathbf{G}\mathbf{G}^H$ correspond to the output signal power and output noise power at $m$-th symbol, and they are shown in Figs. \ref{Fig:signal} and \ref{Fig:noise}, respectively. The ratio of Fig. \ref{Fig:signal} and Fig. \ref{Fig:noise} yields the output SNR with respect to symbol index $m$ and is shown in Fig. \ref{Fig:output_snr}. Following the similar derivations, the output signal power, output noise power, and output SNR are plotted for other waveforms in Fig. \ref{Fig.output_snr_all}.

First, according to Fig. 3, unlike (chirped) DFT-s-OFDM, AFDM, and OTFS, the output signal power, output noise power, and output SNR of OFDM differ for some symbols. This is because OFDM symbols are modulated in the frequency domain and different symbol would experience different fading due to multipath channel. For other waveforms, each symbol is transmitted across nearly the entire frequency band, enhancing their resilience to frequency-selective fading. Second, the proposed chirped DFT-s-OFDM produces lower output noise power and higher output SNR than DFT-s-OFDM. This is due to the inclusion of chirping, facilitating full-band transmission. For example, in Fig. 1b, with $\mathrm{SF}=4$, $2$ unknown data symbols can be estimated from the received data symbols at $8$ subcarriers for chirped DFT-s-OFDM rather than $2$ subcarriers for DFT-s-OFDM. There are more observations than unknowns and the redundant information from more equations can mitigate the impact of noise. Third, the chirped DFT-s-OFDM outperforms AFDM and OTFS in terms of output noise power and output SNR. This is because the introduction of DFT precoding in chirped DFT-s-OFDM spreads data symbols in the time domain. For example, in Fig. 1b, the data symbols are transmitted two times or four times over time, and the retransmission could reduce the impact of noise.

\section{Simulation Results}
The performance of chirped DFT-s-OFDM is evaluated and compared with the existing waveforms in single-user uplink MAC, including DFT-s-OFDM, AFDM, OTFS, and OFDM. {To ensure a fair comparison, different waveforms are compared with identical bit to noise ratio Eb/N0. The energy per (active) subcarrier for (chirped) DFT-s-OFDM remains the same to that for OFDM, AFDM, and \mbox{OTFS}.} Unless otherwise stated, the values of simulation parameters are set as follows. The IFFT size of (chirped) DFT-s-OFDM, AFDM, and OFDM is $N=256$. The DFT size of (chirped) DFT-s-OFDM is $M=64$, leading to the DFT spreading of $\rm{SF}=4$. The chirp rate is $c_{\rm{r}}=1/N$. The carrier frequency and subcarrier spacing are $f_{\rm{c}}=4\,\rm{GHz}$ and $\Delta f=15\,\rm{KHz}$. The $3$-path \mbox{equal-gain} channel ($L=3$) is assumed. The velocity is $v=500\,\rm{km/h}$, corresponding to Doppler frequency of $f_{\rm{max}}=2\,\rm{KHz}$. {For each Monte Carlo simulation, the values of Doppler shifts are randomly generated from $-f_{\rm{max}}$ to $f_{\rm{max}}$.} For OTFS, the numbers of delay grids and Doppler grids are $M_{\rm{OTFS}}=16$ and $N_{\rm{OTFS}}=16$. Quadrature phase shift keying~(QPSK) modulation and LMMSE equalizer are adopted.

 Fig. \ref{Fig.papr_ber_snr}a shows the complementary cumulative distribution function (CCDF) of PAPRs of proposed chirped DFT-s-OFDM and existing waveforms. The proposed waveform by multiplying chirp signal with DFT-s-OFDM preserves the lower PAPR of DFT-s-OFDM over AFDM/OFDM and OTFS. {Note that (chirped) DFT-s-OFDM in Fig. \ref{Fig.papr_ber_snr}a has a PAPR with $0\,\rm{dB}$. This is because DFT-s-OFDM symbol after interleaved subcarrier mapping is the original un-modulated symbols with reduced amplitude and its PAPR remains identical to that of un-modulated symbols \mbox{\cite{4099344,9143507,4085722}}. When phase shift keying (PSK) modulation is employed, a PAPR of $0\,\rm{dB}$ can thus be achieved.}

 \begin{figure}[htbp]
	\centering
		\subfloat[]
		{
			\includegraphics[width = 0.3\textwidth]{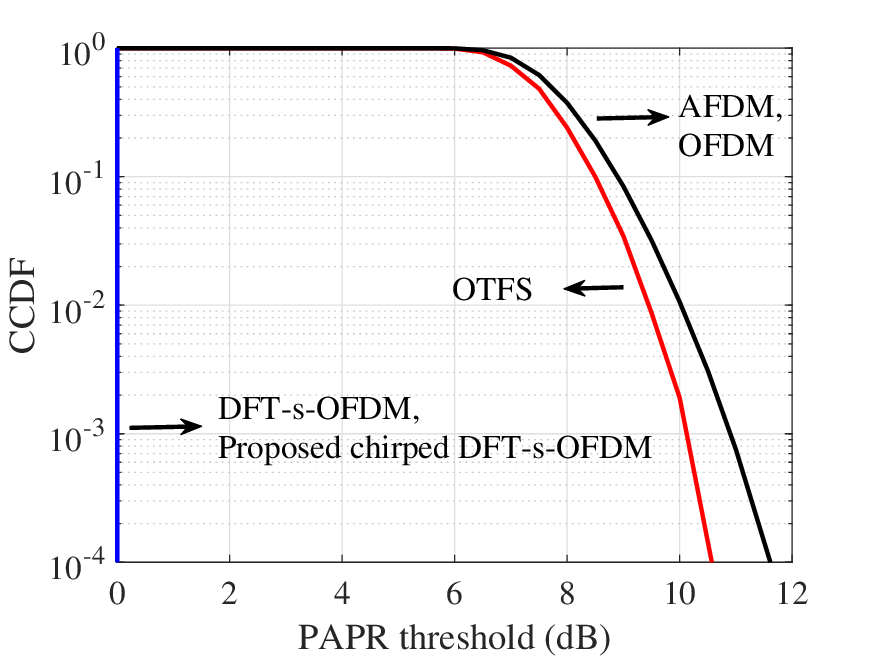}}	
\vspace{0.001cm}
		{
			\includegraphics[width = 0.5\textwidth]{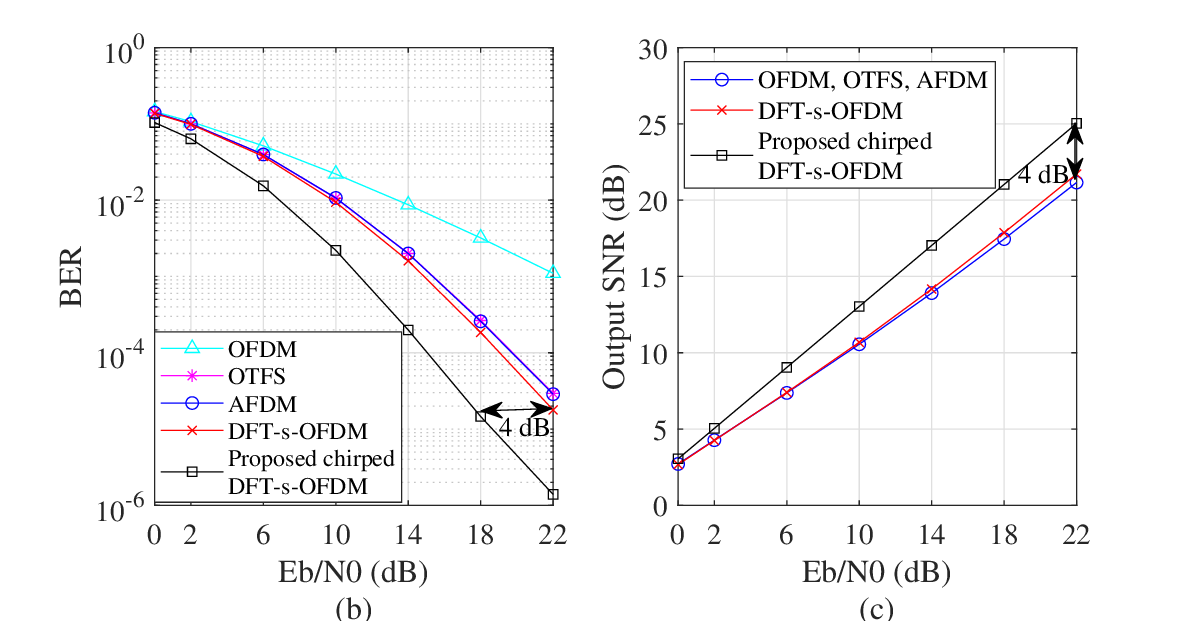}}			
\caption{a) PAPR, b), BER , and c) output SNR of proposed chirped DFT-s-OFDM, in comparison with existing waveforms.}
	\label{Fig.papr_ber_snr}
\end{figure}

Figs. \ref{Fig.papr_ber_snr}b and \ref{Fig.papr_ber_snr}c present the BER and output SNR of proposed chirped DFT-s-OFDM, compared to DFT-s-OFDM, AFDM, OTFS, and OFDM. Thanks to enhanced noise suppression discussed in Subsections II-B and III-B, the proposed chirped DFT-s-OFDM outperforms the existing waveforms in terms of BER and output SNR, with around $4\,\rm{dB}$ $\rm{SNR}$ gain over AFDM, OTFS, and DFT-s-OFDM. Though OFDM has output SNR that is as good as that of OTFS and AFDM, it has deep fades as shown in Fig. \ref{Fig.output_snr_all}. Meanwhile, it cannot exploit frequency diversity, and thus, its BER is worst in Fig. \ref{Fig.papr_ber_snr}b).

Fig. \ref{Fig.ber_output_snr_sf}a) studies the impact of DFT spreading factor $\mathrm{SF}$ on BER of (chirped) DFT-s-OFDM. The BER of DFT-s-OFDM would not obviously enhance with $\mathrm{SF}$. For the proposed chirped DFT-s-OFDM, there is an obvious BER improvement especially when $\mathrm{SF}$ increases from $2$ to $4$. This is because the increase of $\mathrm{SF}$ tends to enhance noise suppression  as can be seen in Fig. \ref{Fig.ber_output_snr_sf}b). As discussed in Subsection II-B, the enhanced noise suppression results from full band transmission and symbols retransmission enabled by chirping and DFT precoding, respectively. {After $\mathrm{SF}=4$, the performance enhancement of chirped DFT-s-OFDM is not obvious. This is because the increase of $\mathrm{SF}$ would reduce signal amplitude by $1/\sqrt{\mathrm{SF}}$ \cite{4099344,9143507,4085722}. The amplitude reduction is significant as $0.35$ and $0.25$ for $\mathrm{SF}=8$ and $\mathrm{SF}=16$, respectively. This amplitude reduction would result in performance degradation, cancelling out the additional benefits gained from additional repeated transmission. Thus, when $\mathrm{SF}$ further exceeds, there is no substantial BER improvement.}

\begin{figure}[htbp]
	\centering
		\includegraphics[width = 0.5\textwidth]{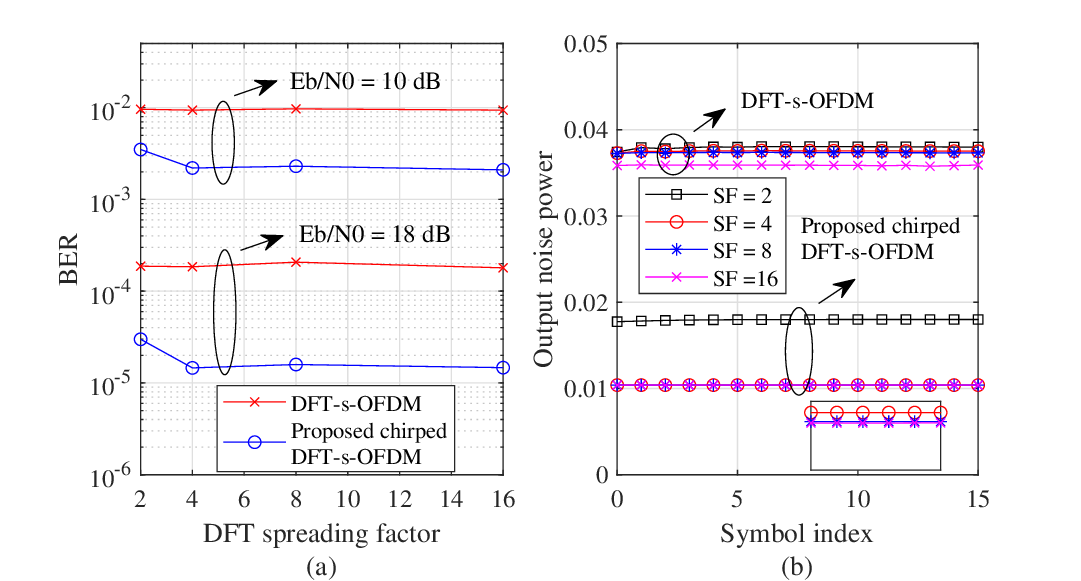}		
\caption{Impact of DFT spreading factor $\mathrm{SF}$ on a) BER and b) output noise power of (chirped) DFT-s-OFDM.}
	\label{Fig.ber_output_snr_sf}
\end{figure}

{Note that, in the previous results, (chirped) DFT-s-OFDM with $\mathrm{SF}>1$ exhibits a different spectral efficiency compared to AFDM and OTFS using all subcarriers. Fig. \ref{Fig.ber_clipping}a shows the BERs of different modulation waveforms under the same bandwidth efficiency with $\mathrm{SF}=4$. AFDM and OTFS use a similar data mapping scheme as (chirped) DFT-s-OFDM, with one-fourth of the interleaved subcarriers (for AFDM) or delay grids (for OTFS) allocated to a specific user for data transmission. Note that the unused subcarriers or delay grids can be assigned to other users for data transmission. Compared to Fig. \ref{Fig.papr_ber_snr}b, the BERs of AFDM and OTFS in Fig. \ref{Fig.ber_clipping}a are lower and are nearly identical to that of chirped DFT-s-OFDM. This implies that the superior BER of chirped DFT-s-OFDM over AFDM and OTFS in Figs. 3 and 4 mainly results from DFT precoding. Though AFDM, OTFS, and chirped DFT-s-OFDM exhibit similar BER in Fig. \ref{Fig.ber_clipping}a, the first two produce higher PAPR than the third. In practical implementation, when the input signal is too high, the power amplifier's output would reach its maximum limit and is unable to increase further, resulting in clipping \cite{9088986,9345729}. When the clipping ratio is set to $1$ \cite{9345729}, the BERs of different waveforms under the same bandwidth efficiency are simulated and shown in Fig. \ref{Fig.ber_clipping}b. AFDM and OTFS are found to be more sensitive to clipping and their BERs are worse than (chirped) DFT-s-OFDM. This is because the signals with high PAPRs, \emph{e.g.,} AFDM and OTFS, are easily affected by clipping. Besides, considering the same bandwidth efficiency using $\mathrm{SF}=4$, OTFS presents higher PAPR than AFDM and that is why its BER degrades more severely than that of AFDM in Fig. \ref{Fig.ber_clipping}b.}

\begin{figure}[htbp]
	\centering
	\includegraphics[width=0.5\textwidth]{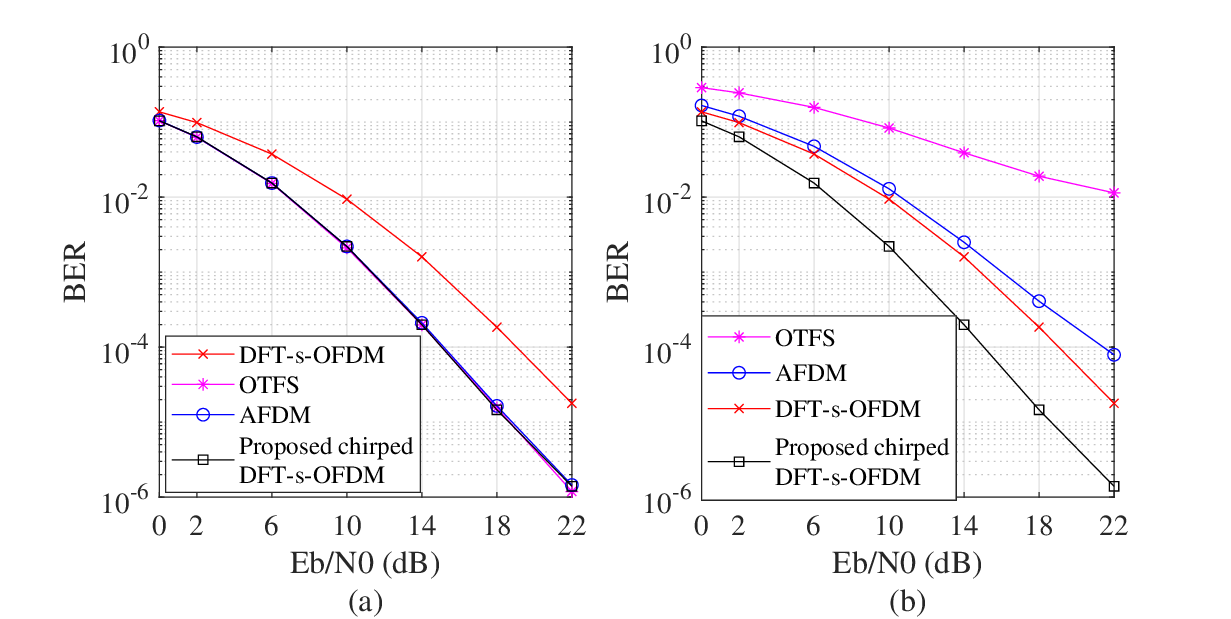}
	\caption{BERs of different modulation waveforms under the same bandwidth efficiency: a) without clipping and b) with clipping ratio of $1$ \cite{9345729}.}
	\label{Fig.ber_clipping}
\end{figure}

\section{Conclusion}
A new single-carrier waveform, called chirped DFT-s-OFDM, has been proposed for 6G communications, with its performance analyzed and evaluated over delay-Doppler channels with \emph{i.i.d.} path amplitudes and different integer delays in single-user uplink MAC in this paper. The new waveform by chirping DFT-s-OFDM in the time domain maintains the low PAPR of DFT-s-OFDM. Thanks to the full band transmission and symbols retransmission enabled by chirping and DFT precoding, respectively, the proposed waveform enables enhanced noise suppression of LMMSE equalization. It presents higher output SNR and lower BER than DFT-s-OFDM, AFDM, OTFS, and OFDM in single-user uplink MAC. The PEP analysis and simulation results confirm that the proposed chirped DFT-s-OFDM can achieve full frequency diversity and demonstrate higher resilience to Doppler shifts than DFT-s-OFDM over delay-Doppler channels with \emph{i.i.d.} path amplitudes and different integer delays.%

Note that the proposed chirped DFT-s-OFDM is applicable to multi-user uplink and downlink MAC. Multiple access interference can be mitigated using an ML equalizer to preserve the performance advantage of proposed waveform. In multi-user downlink MAC, the PAPR of (chirped) DFT-s-OFDM would increase but could be still lower than that of AFDM and OFDM. This is because its PAPR is mainly dependent on the value of spreading factor rather than the number of total subcarriers. Since ML equalizer is computationally inefficient and LMMSE equalizer is likely to be susceptible to multiple access interference, future work aims to develop new multi-user receivers for chirped DFT-s-OFDM to eliminate multiple access interference while with low complexity.

\bibliographystyle{IEEEtran}
{\small
\bibliography{references}}
\addcontentsline{toc}{section}{References}

\ifCLASSOPTIONcaptionsoff
  \newpage
\fi

\end{document}